\documentclass[10pt,conference,a4paper]{IEEEtran}

\usepackage{booktabs}
\usepackage{tabulary}
\usepackage{tabularx}
\usepackage{multirow}
\usepackage{graphicx}
\usepackage{caption}
\usepackage{color}
\usepackage{amsmath}
\usepackage{amssymb}
\usepackage[tight,footnotesize]{subfigure}
\usepackage{todonotes}

\usepackage{fancyhdr}
\fancypagestyle{empty}{
\fancyhf{}

\lfoot{\parbox[]{0.95\textwidth}{\MakeLowercase{
\MakeUppercase{T}his article has been accepted in \MakeUppercase{5GA}rch 2017: 4th \MakeUppercase{I}nternational \MakeUppercase{W}orkshop on \MakeUppercase{5G} \MakeUppercase{A}rchitecture (\MakeUppercase{IEEE ICC' 17}). \MakeUppercase{Preprint Version}.
}}}}
\begin{document}
\bstctlcite{IEEEexample:BSTcontrol}






%

\title{Analytical Modeling for Virtualized Network Functions}
\author{\IEEEauthorblockN{Jonathan Prados-Garzon, Pablo Ameigeiras, Juan J. Ramos-Munoz, Pilar Andres-Maldonado, Juan M. Lopez-Soler}
\IEEEauthorblockA{Department of Signal Theory, Telematics and Communications \\
University of Granada\\
Granada, Spain\\
Emails: jpg@ugr.es, pameigeiras@ugr.es, jjramos@ugr.es, pam91@correo.ugr.es, juanma@ugr.es}}
\maketitle

\date{14 October 2016}

\maketitle
\begin{abstract}

Network Function Virtualization (NFV) is considered one of the key technologies for the 5G mobile networks. In NFV, network functions are implemented in software components denominated Virtual Network Functions (VNFs) running on commodity hardware. 
In this paper, we propose an analytical model based on an open queuing network of G/G/m queues to model VNFs with several components, and chains of VNFs. 
Our model is flexible and generic enough to capture the behavior of such systems. We validate our model by simulation. Specifically,
we validate it for an LTE virtualized Mobility Management Entity with a three-tiered architecture use case.
We also compare our model with 
the estate of the art, in terms of computational complexity and estimation error.
The results show that our model has a computational complexity 
similar to the method for analyzing Jackson's networks. Additionally, our model exhibits an estimation error, measured as the relative error for the estimation of the mean response time, approximately equal to 10\%, whereas for the considered baseline systems it ranges roughly from 60\% to 90\%.     
\end{abstract}

%
%


%
%

%
%

\begin{IEEEkeywords}
5G, NFV, VNF, analytical model, queuing model.
\end{IEEEkeywords}

\section{Introduction}
\label{sec:introduction}


Fifth Generation (5G) mobile networks are expected to be an unprecedented revolution in broadband wireless communications. 
Compared to its predecessor, the Long Term Evolution (LTE) technology,  demanding requirements for 5G 
systems include, among many others, higher network flexibility and scalability, as well as x100 increase in cost effectiveness and energy efficiency \cite{imt2020}.

These  requisites 
compel to revisit the design of the whole system, that is to re-architect the mobile network. In this sense, paradigms such as Network Function Virtualization (NFV) and Software Defined Networking (SDN) 
 are envisaged as cornerstones to build the next 5G mobile network generation.
 
Network Functions (NFs) are functional blocks within a network infrastructure that has well-defined external interfaces and well-defined functional behavior \cite{nfv-use-cases}. Traditionally, most of the NFs have been implemented on 
special-purpose hardware with strongly coupled vendor-dependent software; NFV 
breaks this approach by
disjointing software components from vendor-dependent hardware.

Under the NFV approach, commodity hardware executes software components, so called as Virtual Network Functions (VNF), to provide network functionality.
NFV technology entails numerous advantages, among others: i) lower costs, ii) greater flexibility and scalability, iii) easier and accelerated deployment of new services. 

In NFV, multiple VNFs might be connected to form a service chain. For example, the virtualization of Evolved Packet Core (EPC) and IP Multimedia Subsystem (IMS) is one of the major use cases of NFV \cite{nfv-use-cases}, \cite{Hawilo201418}. And each VNF of a service chain, in turn, might consist of one or more components (VNFCs) working together. A VNFC provides part of the VNF's functionality. Moreover, while a VNF has standardized external interfaces, it admits varying internal designs and operations. 
This yields complex systems with many design options, where it is necessary an agile method for analyzing its performance. As stated in \cite{bhamare16}, there is a need for analytical models for such systems.
For this goal, this paper proposes a generic mathematical model based on an open queuing network for VNFs with a single or multiple VNFCs, which can be easily extended for chains of VNFs. This model provides closed-form expressions to estimate the system response time. 
As discussed in \cite{Urgaonkar05} and \cite{pradosTVT}, such a model has the following applications: 
i) dynamic resource provisioning, which enables a system to adapt its computational resources autonomously depending on the current workload; ii) performance prediction, which provides a performance metric for a given workload and a VNF setup; iii) identifying bottlenecks for purposes of tuning; iv) request policing, which allows VNFs to decline excess requests during temporary overloads.

In our model, each queue represents a VNFC of a VNF running on a VM. Please note that in this work we do not consider the containerization which is an OS-level virtualization method. Each queue in the model might be attended by multiple severs, each one stands for an instance of the computational resource acting as the bottleneck of the VM 
working in parallel (e.g., CPU cores or I/O disks). 

Furthermore, we make no assumptions on the distribution of the packet arrival process or service time at any queue. It is assumed probabilistic routing, where the interconnection between queues is specified by a routing matrix containing the transitions probabilities. This approach is general enough to capture any packet flow through VMs. 

To analyze the queuing network, we adapt the methodology proposed in \cite{whitt83} which is an approximate method to derive the performance metrics of a network of G/G/m queues.

The main advantages of our approach are the following: i) its computational complexity depends on the number of queues $K$, $O(K)$, ii) it provides exact results for Jackson's networks, iii) it is easy to use and employs a reduced set of input parameters. Moreover, our method considers a parameter for each queue to capture the creation or combination of packets at the nodes. This parameter can be used, for instance, to model packet dropping in a virtualized firewall. 

As a result,  we provide simple closed-form expressions to estimate the mean response time of the system with a reduced set of input parameters (see Table \ref{tab:model-input-parameters}). 

We validate our model by simulation. Specifically, we 
consider a particular use case of a virtualized Mobility Management Entity (vMME) following a three-tiered architectural option, i.e., it has three VNFCs. 
The operation assumed for the vMME is similar to that one presented in \cite{Takano14}. For that purpose, we developed a vMME simulator in ns-3 environment that includes the interactions with other EPC entities (e.g., S-GW and eNB). In the evaluation, we generate the signaling workload at the vMME according to the compound traffic model of \cite{pradosTVT}.

The 
estimation error of the proposed model is compared to baseline schemes. Specifically, results show that the relative error for the estimation of the mean response time of our model for the considered use case is approximately equal to 10\%, whereas the estimation error of baseline methods ranges roughly from 60\% to 90\%.  

The remainder of the paper is organized as follows. Section \ref{sec:background-and-related-works} provides some background and briefly describe related works. Section \ref{sec:system-and-operation-model} describes the system model. In Section \ref{sec:a-model-for-a-three-tier-vne} we propose the queuing model for a VNF design including system response time expressions. In Section \ref{sec:use-case_a-three-tier-vmme} we particularize our model to a specific vMME use case. Section \ref{sec:numerical-results} includes numerical results for model validation. Finally, Section \ref{sec:conclusions} draws the main conclusions.

\section{Background and Related Works}
\label{sec:background-and-related-works}

This section briefly reviews the proposed designs for virtualizing NFs and revises the related literature. 


\subsection{VNF desings}
\label{subsec:scalable-vnf-design}

The design of the VNFs is substantial when exploiting all the advantages NFV offers. Accordingly, the 1:N mapping (or multi-tiered) architecture, inspired by the web services designs, has been proposed for VNFs due to its robustness, flexibility and scalability \cite{Taleb15}. In such design every network function is split into multiple VNFCs -or tiers- of the following three types \cite{Taleb15}: i) Front-End (FE), which might act as an interface to the outside world and load balancer; ii)  Worker (W), which implements the logic of the network function; and iii) state DataBase (DB) that contains the state information.

There are several works that consider this architecture to virtualize EPC entities such as MME \cite{Takano14}, \cite{premsankar15}, \cite{pradosTVT}. It is also be applied to virtualize the IMS \cite{carella14}.

In the opposite direction, N:1 mapping design groups several NFs into one virtual component. This design minimizes the number of interfaces and offers a highly low-delay processing \cite{Taleb15}. In \cite{baba15}, the authors propose the use of this architectural option to deploy lightweight virtualized EPCs tailored for each particular Machine-to-machine service.

\subsection{Related works}
\label{subsec:related-works}

In \cite{Urgaonkar05}, the authors propose a closed queuing network tailored to model Multi-tier Internet Services. This substantial work has served as a basis for many other works. 
To calculate the mean response time, the authors use the iterative algorithm Mean Value Analysis (MVA). Therefore, no closed-form expressions are provided to compute performance metrics. 
And their proposed MVA algorithm has a time complexity depending on the number of active sessions. 
In mobile networks scenarios, typically characterized by a large number of active sessions, MVA execution might be longer than the simulation of the scenario, as shown in Sec. \ref{sec:numerical-results}.  

In \cite{Urgaonkar08} and \cite{bi10}, the authors address the dynamic resource provisioning problem. For that, they employ open queuing networks to model multi-tier applications. In \cite{Urgaonkar08}, each tier is modeled independently as a G/G/1 queue, but in this case, the model does not capture the operation of the system. Moreover, they do not provide a method to compute the internal flows parameters from the external arrival process. In \cite{bi10}, the authors assume an M/M/m queue to model the first tier and M/M/1 queues for the rest of the tiers. This model offers limited flexibility because it considers a Jackson's network and, consequently, the arrival and service processes are fixed.


In our previous work \cite{pradosTVT}, we propose a model for vMME dimensioning. That work also provides a scalability analysis of a vMME. To that end, we used a model based on an open Jackson Network for a vMME with a 1:3 mapping architecture. We show that the  proposed vMME model provides fairly good results for resource dimensioning. 
In this work, we enhance that model by extending its generality in the following points:
\begin{itemize}
\item We extend its scope of application to chains of VNFs, and VNFs with multiple VNFCs and an arbitrary design.
\item We improve its capacity to capture different behaviors of the system (e.g., feedback, generic load balancers, caching effects at any stage, hardware heterogeneity,...).  
\end{itemize}

\section{System Model}
\label{sec:system-and-operation-model}

\begin{figure}[t]
\begin{center}
\includegraphics[width=0.7\columnwidth,height=5cm]
{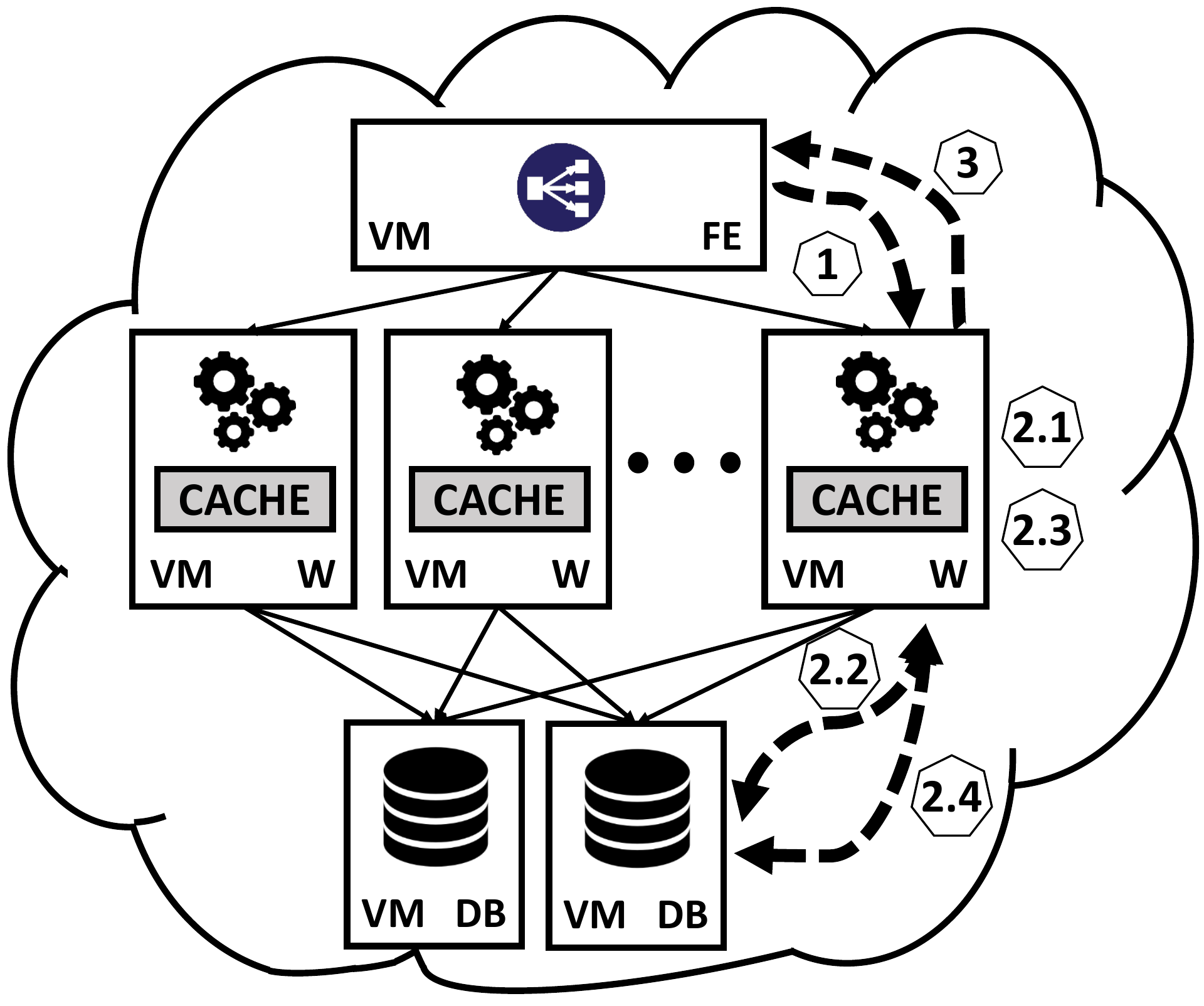}
\end{center}
\caption{System model and operation.}
\label{fig:system-model-and-operation}
\end{figure}

In this work, without loss of generality, we consider a VNF with a 1:3 mapping architecture (see Fig. \ref{fig:system-model-and-operation}). Specifically, the VNF is decomposed into three VNFCs: the Front-End (FE) ii) Workers (Ws), and iii) the state DataBase (DB). The FE is the communication interface with other network entities and balances the load among the Ws. Each worker implements the logic of the VNF functionality, and the DB contains the user session state making the Ws stateless. 

Let us now describe the general operation of a VNF with a 1:3 mapping design, whose main steps are depicted in Fig. \ref{fig:system-model-and-operation}.
Since the FE acts as the communication interface with the outside world, all packets enter the VNF at the FE with a mean rate $\lambda_{0FE}$. Then, the FE sends the packet to the corresponding W according to its load-balancing scheme (labeled as "1"). Once the packet arrives at the W, the W parses the packet and checks whether the required data for processing the packet are stored in its cache memory (labeled as "2.1"). This cache memory could be implemented inside the RAM allocated to the Virtual Machine (VM), where the W is running on. If a cache mismatch occurs, then the W forwards a query to the DB to retrieve the data from it (labeled as "2.2"). Please note that this data retrieval leads a preemption in the packet processing at the W, during which the W might process other packets. When the W has the necessary data, it immediately keeps processing the packet (labeled as "2.3"). After processing finishes, it might be necessary to update some data in the DB (labeled as "2.4"). Then, the W generates a response packet and forwards it to the FE (labeled as "3"). Finally, the packet exits the VNF. 

We assume that the VNF is run in a data center. We also consider that all VMs, hosted on the same physical machine, do not share any computational resource. Then, any VM does not experiment dynamic changes in performance at runtime (no-DCR assumption) \cite{DynamicCloudSim}.  

In general, each VNFC (e.g., FE, W, and DB) can scale horizontally and might, therefore, have several instances deployed in separate VMs. Two different VMs might offer distinct performance, even if they implement the same element. Reasons may be because a given VM was scaled vertically, which is achieved through live migration in current clouds, or due to the heterogeneity of hardware in the data center. 

It is assumed that every element instance serves the packets following a First-Come, First-Served (FCFS) discipline. The service rate $\mu$ for an instance is given by its computational resource acting as the bottleneck. The tasks on VMs, mostly, might be CPU-, I/O-, or communication-intensive \cite{DynamicCloudSim}. For instance, the Ws are CPU-bound,
whereas DB applications are primarily I/O-bound. Let $NI$ be the number of instructions to be executed for a given task, and $r_{CPU}$ the CPU power. The execution time of the task, under no-DCR assumption, can be estimated as $t_{e} = NI/r_{CPU}$. Similarly, the task execution time can be estimated for an I/O-bound task from its number of I/O operations and the performance of the VM disk (measured as I/O operations per second -$IOPS$-).  

By way of example, let us consider a W instance, where the processing of the packets is CPU-intensive. And let us suppose there are $H$ different types of packets, each one with an associated frequency of occurrence $f^{(H)}$. And a W executes $NI^{H}$ instructions to process each packet type. Then, the mean and the variance of the W's service time, denoted as $s=1/\mu$ and $\sigma_{s}^2$, can be estimated as:
\begin{equation}
{s = \sum_{H}f^{(H)}\frac{NI^{(H)}}{r_{cpu}}}
\label{eq:mean-service-time-w}
\end{equation}
\begin{equation}
{\sigma_{s}^2 = \sum_{H}f^{(H)}\left(\frac{NI^{(H)}}{r_{cpu}}-s\right)^2}
\label{eq:var-service-time-w}
\end{equation}

Finally, the VNF has to meet a given performance requirement. The performance requisite considered in this work is a bound on the mean response time that is acceptable to the VNF. The VNF can adapt its capacity as a function of the workload. For instance, when the processing capacity assigned to the VNF cannot withstand with the current workload, it will be necessary to increase the capacity of the VNFC that currently represents the bottleneck of the system. This can be achieved by scaling up some instances of the component (vertical scaling) or by scaling out the component (horizontal scaling). In the same way, when the computational resources allocated to a given element are underused, some instances of the element are scaled down or removed.  


\section{Analytical Model for VNFs}
\label{sec:a-model-for-a-three-tier-vne}
This section explains the queuing model for a VNF and the methodology of analysis considered to derive the system response time from the model.


  \begin{table}[tb]
\centering
\caption{Model input parameters.}
\label{tab:model-input-parameters}
\begin{tabular}{ll}
\hline
\hline
Notation & Description \\
 \hline
 $\lambda_{0k}$ & Mean external arrival rate at queue $Q_{k}$. \\
 $c_{0k}^2$ & SCV of the external arrival process at queue $Q_{k}$.\\
 $m_{k}$ & Number of servers at queue $Q_{k}$.\\
 $\mu_{k}$ & Average service rate at queue $Q_{k}$.\\
  $c_{sk}^2$ & SCV of the service process at queue $Q_{k}$.\\
 $K^{(j)}$ & Number of instances of the $j$th stage.\\
  $P=[p_{ik}]$ & Routing probability matrix.\\
  $\nu_{k}$ & Multiplicative factor for the flow leaving $Q_{k}$.\\
\hline
\hline
\end{tabular}
\end{table}


\subsection{Queuing Model}
\label{subsec:queuing-model}

\begin{figure}[t]
\begin{center}
\includegraphics[width=1.0\columnwidth]{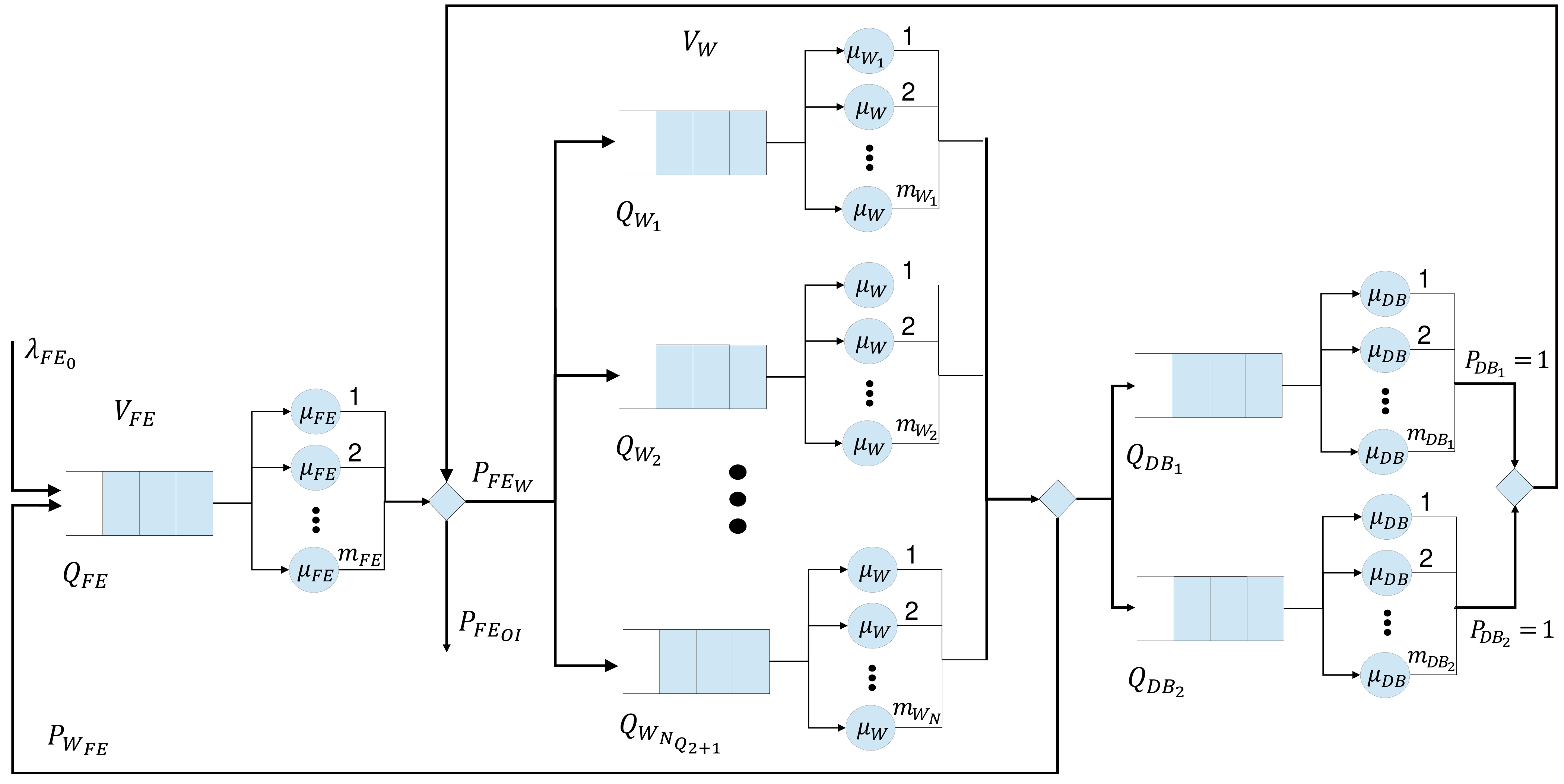}
\end{center}
\caption{Queuing model for a VNF with a three-tier design.}
\label{fig:queuing-model}
\end{figure}

Let us consider a VNF chain with $J$ stages, where a stage is a VNFC. 
To model this system we employ an open network of $K$ G/G/m queues $Q_{1}, Q_{2}, \cdots, Q_{K}$ (see Fig. \ref{fig:queuing-model}). Since every stage can be scaled horizontally, each queue represents an VNFC instance (or replica) running on a VM. 

Let $K_{j}$ denote the number of instances per stage. 
Then, $K=\sum_{j=1}^{J} K_{j}$.
The set of queue indexes $k \in \mathbb{N}$ for a given stage $j \in \mathbb{N} $ verifies that $1 \leq k \leq K_{1}$ if $j=1$, and $\sum_{i=1}^{j-1} K_{i} + 1 \leq k \leq \sum_{i=1}^{j} K_{i}$ if $j \geq 1$.

Moreover, each queue might have $m_{k}$ servers which may represent, for instance, different CPU cores processing messages in parallel.

Every queue is assumed to operate under Fist-Come, First-Served (FCFS) discipline.

Regarding the external arrival process to each queue $Q_{k}$, it is assumed to be a generalized inter-arrival process, which is characterized by its mean $\lambda_{0k}$ and its squared coefficient of variation ($SCV = variance/(mean)^2$) $c_{0k}^{2}$.

We consider that all servers of the same queue have an identical and generalized service process, which is also characterized by its mean $\mu_{k}$ (service rate) and its SCV $c_{sk}^{2}$. However, servers belonging to different queues may have distinct service processes, even if they pertain to the same tier. This feature is useful to model the heterogeneity of the physical hardware, underlying the provisioned VMs, inherent to non-uniform infrastructures like computational clouds \cite{DynamicCloudSim}.

Furthermore, every queue has associated a parameter $\nu_{k}$, which is a multiplicative factor for the flow leaving $Q_{k}$ that models the creation or combination of packets at the nodes. That means that if the total arrival rate to queue $Q_{k}$ is $\lambda_{k}$, then the output rate of this queue would be $\nu_{k} \lambda_{k}$. 

For the transitions between queues, we assume probabilistic routing where the packet leaving $Q_{k}$ is next moved to queue $Q_{i}$ with probability $p_{ki}$ or exits the network with probability $p_{0k}=1-\sum_{i=1}^{K} p_{ki}$. We also assume that the routing decision is made independently for each packet leaving queue $Q_{k}$.

The transition probabilities $p_{ki}$ are gathered in the routing matrix denoted as $P=\left[ p_{ki}\right]$. This approach allows to define any arbitrary feedback between stages, and to model caching effects and different load-balancing strategies at any stage.


\subsection{System Response Time}
\label{subsec:system-response-time}

To compute the system response time, we use the approximation technique proposed in \cite{whitt83} for the Queueing Network Analyzer (QNA). This technique, hereinafter referred as the QNA method, uses two parameters, the mean and the SCV, to characterize the arrival and service time processes for every queue. Then the different queues are analyzed in isolation as standard GI/G/m queues.

Finally, to compute the global performance parameters, the QNA method assumes that the queues are stochastically independent, even though the queuing network might not have a product-form solution. Thus, QNA method can be seen as a generalization of the open Jackson's network of M/M/m queues to an open Jackson's network of GI/G/m queues \cite{whitt83}.


As we will show in Section \ref{subsec:model-validation}, although the QNA method is approximate, it provides a remarkably good approximation to the values obtained by simulation. We describe the main steps of the QNA method in the following paragraphs.

\subsubsection{Internal Flows Parameters Computation}
\label{subsubsec:internal-flows-parameters-computation}
The first step of the QNA method is to compute the mean and the SCV of the arrival process to each queue.

Let $\lambda_{k}$ denote the total arrival rate to queue $Q_{k}$. As in the case of Jackson's networks, we can compute $\lambda_{k}, \enskip \forall \enskip \{ k \in \mathbb{N}  |  1 \leq k \leq K \}$ by solving the following  set of linear flow balance equations:
\begin{equation}
{\lambda_{k} = \lambda_{0k} + \sum_{i=1}^{K} \lambda_{i} \nu_{i} p_{ik}}
\label{eq:lambdaj}
\end{equation}

Let $c_{ak}^{2}$ be the SCV of the arrival process to each queue $Q_{k}$. To simplify the computation of the $c_{ak}^{2}$, the QNA method employs approximations. Specifically, it uses a convex combination of the asymptotic value of the SCV $(c_{ak}^2)_{A}$ and the SCV of an exponential distribution ($c_{exp}^{2} = 1$), i.e., $c_{ak}^{2}=\alpha_{k} (c_{ak}^2)_{A} + (1-\alpha_{k})$. The asymptotic value can be found as $(c_{ak}^2)_{A} = \sum_{i=1}^{K} q_{ik}c_{ik}^2$, where $q_{ik}$ is the proportion of arrivals to $Q_{k}$ that came from $Q_{i}$. That is, $q_{ik} = \frac{\lambda_{i} \nu_{i} p_{ik}}{\lambda_{k}}$. And $\alpha_{k}$ is a function of the server utilization $\rho_{k} = \frac{\lambda_{k}}{\mu_{k} m_{k}}$ and the arrival rates. This approximation yields the following set of linear equations, which may be solved to get $c_{ak}^{2}, \enskip \forall \enskip \{ k \in \mathbb{N} |  1 \leq k \leq K \}$:
\begin{equation}
{c_{ak}^2 = a_{k} + \sum_{i=1}^{K} c_{ai}^{2} b_{ik}, \qquad 1 \leq k \leq K}
\label{eq:cak}
\end{equation}
\begin{multline}
a_{k} = 1 + \omega_{k} \bigg\{ (q_{0k}c_{0k}^2 - 1) \\ 
+ \sum_{i=1}^{K} q_{ik}[(1-p_{ik}) + \nu_{i} p_{ik} \rho_{i}^{2} x_{i}]\bigg\}
\label{eq:ak}
\end{multline}
\begin{equation}
{b_{ik} = \omega_{k} q_{ik} p_{ik} \nu_{i} (1 - \rho_{i}^{2}) }
\label{eq:bik}
\end{equation}
\begin{equation}
{x_{i} = 1 + m_{i}^{-0.5} (max\{ c_{si}^{2}, 0.2 \} - 1)}
\label{eq:xi}
\end{equation}
\begin{equation}
{\omega_{k} = \left(1 + 4 (1 - \rho_{k})^{2} (\gamma_{k} - 1)\right)^{-1}}
\label{eq:wk}
\end{equation}
\begin{equation}
{\gamma_{k} = \left( \sum_{i=0}^{K} q_{ik}^{2} \right)^{-1}}
\label{eq:gammak}
\end{equation}

\subsubsection{Response Time Computation}
\label{subsubsec:mean-response-time-computation}
Once we have found $\lambda_{k}$ and $c_{ak}^{2}$ for all internal flows, we can compute the performance parameters for each queue, which are analyzed in isolation (i.e., considering that the queues are independent of each other).

Let $W_{k}$ be the mean waiting time at queue $Q_{k}$. If $Q_{k}$ is a GI/G/1 queue ($Q_{k}$ has only one server), $W_{k}$ can be approximated as:
\begin{equation}{W_{k}= \frac{\rho_{k} \cdot (c_{ak}^{2} + c_{sk}^{2}) \cdot \beta}{2 \cdot \mu_{k} (1 - \rho_{k})}}
\label{eq:waiting-time-gg1}
\end{equation}
with
\begin{equation}{\beta = \left\{ 
\begin{array}{ll}
exp ( -\frac{2 \cdot (1 - \rho_{i}) \cdot (1-c_{ai}^2)^2}{3 \cdot \rho_{i} \cdot (c_{ai}^{2} + c_{si}^{2})} )  & \qquad c_{ai}^{2} < 1 \\
\beta = 1 & \qquad c_{ai}^{2} \geq 1
\end{array}
\right.
}
\label{eq:beta}
\end{equation}

If, by contrast, $Q_{k}$ is a GI/G/m queue, $W_{k}$ can be estimated as:
\begin{equation}
{W_{k}= 0.5 \cdot \left(c_{ai}^{2} + c_{si}^{2}\right)\cdot W_{k}^{M/M/m} }
\label{eq:waiting-time-ggm}
\end{equation}
where $W_{k}^{M/M/m}$ is the mean waiting time for a M/M/m queue, which can be computed as:
\begin{equation}
{W_{k}^{M/M/m}=\frac{C(m_{k},\frac{\lambda_{k}}{\mu_{k}})}{m_{k} \mu_{k} - \lambda_{k}}}
\label{eq:waiting-time-mmm}
\end{equation}
and $C(m,\rho)$ represents the Erlang's C formula which has the following expression:
 \begin{equation}
   C(m,\rho)=\frac{\left( \frac{(m\cdot \rho)^m}{m!} \right) \cdot \left( \frac{1}{1 - \rho} \right)}
		  {\sum_{k=0}^{m-1} \frac{(m\cdot \rho)^k}{k!} + \left( \frac{(m\cdot \rho)^m}{m!}\right) \cdot \left( \frac{1}{1-\rho}\right)}
   \label{eq:erlangc}
  \end{equation}

Let $V_{k}$ denote the visit ratio for $Q_{k}$ which is defined as the average number visits to node $Q_{k}$ by a packet during its lifetime in the network. That is, $V_{k}=\frac{\lambda_{k}}{\sum_{i=1}^{K}\lambda_{0k}}$. Finally, the overall mean response time will be given by
\begin{equation}
{ T = \sum_{k=1}^{K} ( W_{k} + \frac{1}{\mu_{k}} ) \cdot V_{k}}
\label{eq:response-time}
\end{equation}

\section{Particular Use Case: a Three-Tier vMME}
\label{sec:use-case_a-three-tier-vmme} 
 In this section, we particularize our model to a vMME with a 1:3 mapping design whose operation is presented in \cite{Takano14}. Specifically, we provide analytical expressions to compute the input parameters of the model for this use case.
 
\subsection{Signaling Workload for the vMME}
\label{subsec:signaling-workload-vmme}
The MME is the main control entity of the LTE/EPC. It interacts with the evolved NodeB (eNB), Serving Gateway (S-GW), and Home Subscriber Server (HSS) within the EPC to realize functions such as non-access stratum (NAS) signaling, user authentication and authorization, mobility management (e.g. paging, user tracking),
and bearer management, among many others \cite{pradosTVT}.

The User Equipments (UEs) run the users' applications that generate or consume network traffic, which, in turn, trigger the LTE network control procedures. These signaling procedures allow the control plane to manage the UE mobility and the data flow between the UE and PDN-GW. These control procedures yield several messages to be processed by the vMME.

Here, we only consider the signaling procedures that generate most signaling load \cite{Hirschman15}, e.g., Service Request (SR), Service Release (SRR), and X2-Based Handover (HR). 

Let $\lambda_{CP}$ and $n_{CP}$ respectively be the average procedure rate per UE and the number of packets to be processed by the MME for each control procedure $P \in \{ SR, SRR, HR \}$. Specifically, $n_{SR} = n_{SRR} = 3$ and $n_{HR} = 2$. Then, we can compute the average number of packets per signaling procedure $\overline{N}_{pp}$ as 
\begin{equation}
{\overline{N}_{pp} = \frac{\sum_{P} \lambda_{P} n_{P}}{\sum_{P} \lambda_{P}} = \frac{3 \lambda_{SR} + 3  \lambda_{SRR} + 2 \lambda_{HR}}{ \lambda_{SR} + \lambda_{SRR} + \lambda_{HR} }}
\label{eq:Npp}
\end{equation}
Let us assume a Poisson arrival process for each control procedure $P$, which is reasonable since we will have a large number of independent UEs $N_{U}$ attached to the LTE network generating signaling workload. 
Then, the overall mean rate $\lambda$ and SCV $c_{a}^{2}$ of the vMME external arrival process are given by:
\begin{equation}
{\lambda = N_{U} \sum_{P} n_{P} \lambda_{P} = N_{U}(3 \lambda_{SR} + 3 \lambda_{SRR} + 2 \lambda_{HR})}
\label{eq:lambda-vmme}
\end{equation}
\begin{equation}
{c_{a}^{2} = 1}
\label{eq:scva-vmme}
\end{equation}

\subsection{Model Particularization for a vMME}
\label{subsec:model-particularization-vmme}

Let us consider a vMME with a 1:3 mapping design as described in Sec. \ref{sec:system-and-operation-model}. Let $K_{FE}$, $K_{W}$, and $K_{DB}$ respectively denote the number of front-end, workers, and database instances. Consequently, $K=K_{FE}+K_{W}+K_{DB}$.

We assume perfect load balancing for all tiers \cite{Urgaonkar05}. That is, each FE, W, and DB instance respectively processes $\frac{1}{K_{FE}}$, $\frac{1}{K_{W}}$, and $\frac{1}{K_{DB}}$ fraction of the total workload of the tier they belong to.

The operation assumed for the vMME is similar to \cite{Takano14} and \cite{premsankar15}. In this use case, the main differences between the assumed vMME behavior and those described in Sec. \ref{sec:system-and-operation-model} are explained bellow.
 
On the one hand, the FE tier balances signaling workload equally among the W instances on a per control procedure basis. The FE sends to the same W instance all control messages associated with a given control procedure and UE. It is assumed that the W instance has enough memory to store all the required state data (e.g., UE context) to handle a control procedure during its lifetime.    

On the other hand, when the initial message of a procedure arrives at any W instance and after the W parses the packet, the W retrieves the UE context from DB. Moreover, when the W instance finishes processing the last message of a signaling procedure, it saves the updated UE context into the DB.

To summarize, there are two DB accesses per control procedure. Therefore, the visit ratio per packet at each DB and W instance will respectively be $V_{DB} = \frac{1}{K_{DB}} \cdot \frac{2}{\overline{N}_{pp}}$ and 
$V_{W}= \frac{1}{K_{W}} \cdot (1 + \frac{2}{\overline{N}_{pp}} )$.

Since the FE maintains 3GPP standardized interfaces towards other entities of the network (e.g., eNBs, HSS, and S-GW), all control messages from outside arrive at the FE tier. 
For the same reason, the vMME sends all response messages generated by W instances out on FE tier. Then, the visit ratio at each FE instance is given by $V_{FE} = \frac{2}{K_{FE}} 
$, i.e., each packet visits the FE tier two times. Furthermore, a packet served at any FE instance leaves the vMME (queuing network) with probability $p_{0FE}=0.5$.




Taking all the above into account and, for simplicity, assuming only one instance for the FE ($K_{FE}=1$) and DB ($K_{DB} = 1$), the routing matrix will be:
\begin{equation}
{ 
P=\left( \begin{array}{ccccc}
0 & \frac{1-p_{0FE}}{K_{W}} & \cdots & \frac{1-p_{0FE}}{K_{W}} & 0 \\
\frac{\frac{\overline{N}_{pp}}{2}}{\frac{\overline{N}_{pp}}{2} + 1} & 0 & \cdots & 0 & \frac{1}{\frac{\overline{N}_{pp}}{2} + 1} \\
\vdots & \vdots & \vdots & \vdots & \vdots \\
\frac{\frac{\overline{N}_{pp}}{2}}{\frac{\overline{N}_{pp}}{2} + 1} & 0 & \cdots & 0 & \frac{1}{\frac{\overline{N}_{pp}}{2} + 1} \\
0 & \frac{1}{K_{W}} & \cdots & \frac{1}{K_{W}} & 0 \\
\end{array} \right)
}
\label{eq:routing-matrix-vmme}
\end{equation}



\section{Numerical Results}
\label{sec:numerical-results}

In this section, we present our experimental setup and numerical results. It aims at validating the described analytical model based on the QNA method \cite{whitt83} of a VNF with a 1:N mapping architecture. In addition, we compare our model with baseline models in terms of accuracy and execution time.

\subsection{Experimental Setup}
\label{subsec:experimental-setup}

To validate the proposed model, we use two software tools both implemented in the ns-3 simulation environment. The first one is the generator of LTE control procedure calls for human-type communications, more details are provided in \cite{pradosTVT}. The traffic model setup and the Radio Access Network (RAN) scenario are also described in the same work. Consequently, Table \ref{tab:parameters-configuration} shows the signaling rates per user.

The second one is a simulator of a vMME behavior described in Sec. \ref{sec:use-case_a-three-tier-vmme} and \ref{sec:system-and-operation-model}, and its interactions (i.e., control messages exchange) with other network entities (e.g. S-GW and eNB). As discussed in \cite{pradosTVT}, the service time at a W instance $t_{W}$ might be different for each control message type. In the simulator, $s_{W}$ is estimated as $s_{W} = \frac{NI^{(H)}}{r_{cpu}}$. 
The interaction of the vMME with S-GW and eNB is simulated as constant delays.

Our analytical model, which uses QNA method, was implemented in GNU Octave.  The external arrival process characteristics was computed by using (\ref{eq:Npp}), (\ref{eq:lambda-vmme}), and (\ref{eq:scva-vmme}). And the routing matrix $P$ is given by (\ref{eq:Npp}) and (\ref{eq:routing-matrix-vmme}) with $p_{0FE}=0.5$.

For the use case considered $\nu_{k}=1 \quad \forall \quad k \in \left[ 1, K\right]$. Since we consider that service rates at FE and DB tiers are deterministic (see Table \ref{tab:parameters-configuration}), $c_{sk}^2=0$ for the corresponding queues.

The mean $s_{W}$ and variance $\sigma_{sW}^2$ of the service time at any W instance was computed by using (\ref{eq:mean-service-time-w}) and (\ref{eq:var-service-time-w}). Then, the SCV of the service process at any W $c_{sW}^2$ is given by $c_{sW}^2 =  \sigma_{sW}^2/s_{W}^2$.

For comparison purposes, we employed the Queueing Network Package for GNU Octave \cite{Marzolla10}. This package includes functions to solve queuing networks by using Jackson's networks assumptions (e.g., qnos function) and MVA algorithm (e.g., qncsmva function). It shall be noted that the input parameters of our model can also be used to these other methodologies for analyzing queuing networks. 

All the experiments were carried out on a machine with 4 GB RAM and Intel Xeon CPU X3330 @ 2.66GHz x 4 processor. The machine ran Linux Ubuntu 12.04 LTS operating system. 

\newcolumntype{P}[1]{>{\centering\arraybackslash}p{#1}}
\begin{table}[tb]
\centering
\caption{Parameters Configuration}
\label{tab:parameters-configuration}
\begin{tabular}{|p{3.9cm}|P{4cm}|}
\hline
\multicolumn{2}{|c|}{Signaling Rates per UE}  \\ \hline\hline 
$\lambda_{SR}$ (Poisson) & 0.0045 procedures/second   \\ \hline
$\lambda_{SRR}$ (Poisson) & 0.0045 procedures/second   \\ \hline
$\lambda_{HR}$ (Poisson) & 0.0012 procedures/second   \\ \hline \hline
\multicolumn{2}{|c|}{EPC Delays}  \\ \hline \hline
One-way delay (eNB $\rightarrow$ vMME) & 4.5 ms \\ \hline
Two-way delay (vMME $\rightleftharpoons$ [eNB $|$ S-GW]) & 9 ms \\ \hline
\multicolumn{2}{|c|}{Service Rates}                        \\ \hline \hline
FE service rate ($\mu_{FE}$)   & 120000 packets per second 
\\ \hline
DB service rate ($\mu_{SDB}$)   & 100000 transactions per second 
\\ \hline
W CPU power ($r_{cpu}$)   & $11.38 \cdot GFLOPS$ 
\\ \hline
($NI^{(H)}$ $H$ $\in$ $\{SR_{1}$, $SR_{2}$, $SR_{3}$, $SRR_{1}$, $SRR_{2}$, $SRR_{3}$, $HR_{1}$, $HR_{2}\}$ \cite{pradosTVT}   & (1.45, 1.07, 1.06, 1.07, 1.07, 1.06, 1.07, 1.07) $\cdot 10^{6} Instructions$ 
\\ \hline
\end{tabular}
\end{table}


\subsection{Execution Time Comparison}
\label{subsec:execution-time-comparison}

\begin{table}[tb]
\centering
\caption{Execution times for solving the queuing network that models a vMME.}
\label{tab:execution-times}
\begin{tabular}{|c|c|c|c|c|c|}
\hline
$N_{U}$ & $K$ & Jackson & QNA & MVA & Simulation \\
\hline
\hline
100000 & 3 & 6.667 ms & 8.000 ms & 9.740 s & 224.280 s\\
\hline
500000 & 4 & 6.000 ms & 9.340 ms & 48.691 s & 241.100 s\\
\hline
1000000 & 5 & 5.333 ms & 10.667 ms & 97.162 s & 252.880 s\\
\hline
2000000 & 8 & 7.000 ms & 21.335 ms & 194.884 s & 251.660 s\\
\hline
\end{tabular}
\end{table}


We assessed the time complexity of our analytical model and compared it with Jackson, MVA, and simulation approaches (see Table \ref{tab:execution-times}). We repeated each measurement three times and average it. 

As can be observed, our model has a time complexity that depends on the number of queues in the network $K$, $O(K)$. While the model based on MVA algorithm \cite{Urgaonkar05} exhibits a time complexity that is a linear function of the number of active user sessions $N_{U}$.

Since $N_{U}$ may be high in cellular scenarios, the MVA algorithm might even take longer than the simulation of the scenario. For all the simulations, the system has to process $4 \cdot 10^6$ packets as the stop condition. We checked that this achieves simulation convergence. The chosen stop condition explains that the simulation execution time does not depend strongly on $N_{U}$.



\subsection{Model Validation}
\label{subsec:model-validation}


\begin{figure}
\centering

\includegraphics[width=\columnwidth]{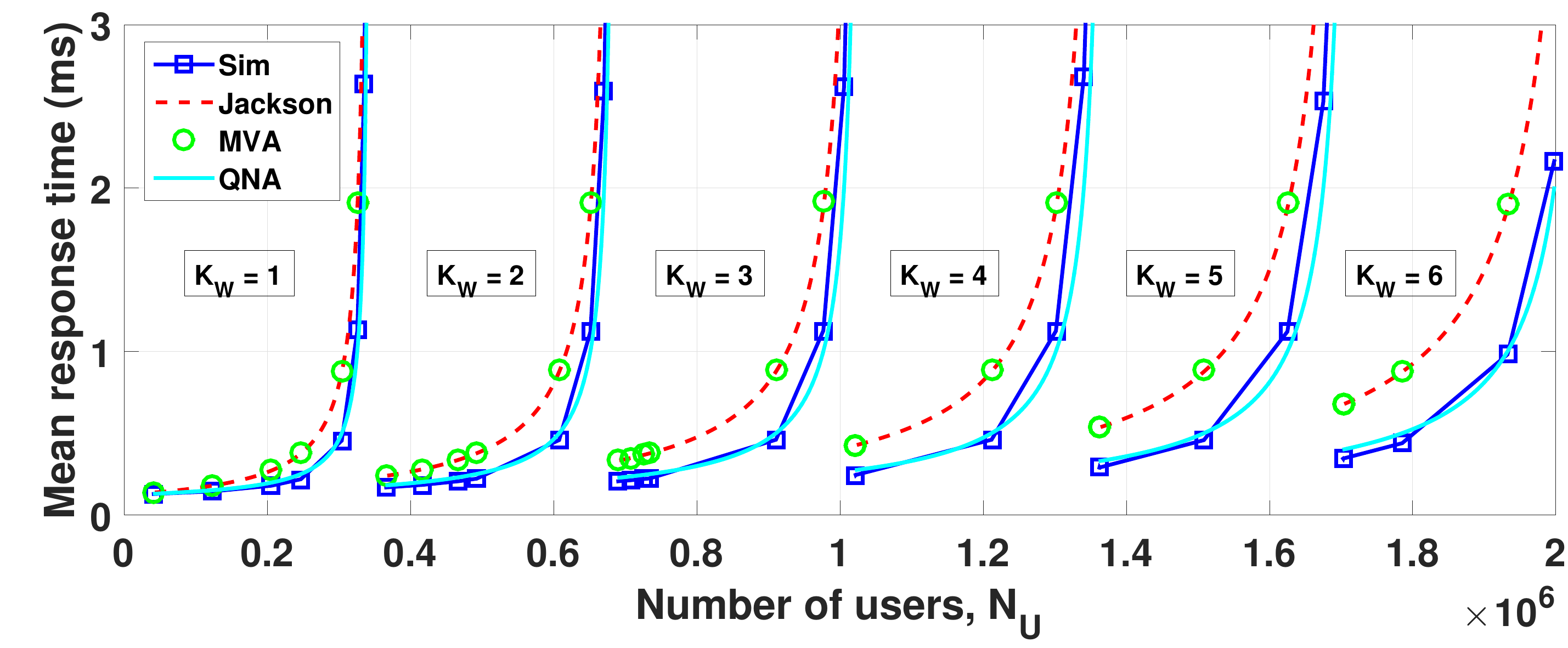}
	\caption{Mean response time.}
	\label{subfig:mean-response-time}
\end{figure}

\begin{figure}
\centering
\includegraphics[width=\columnwidth]{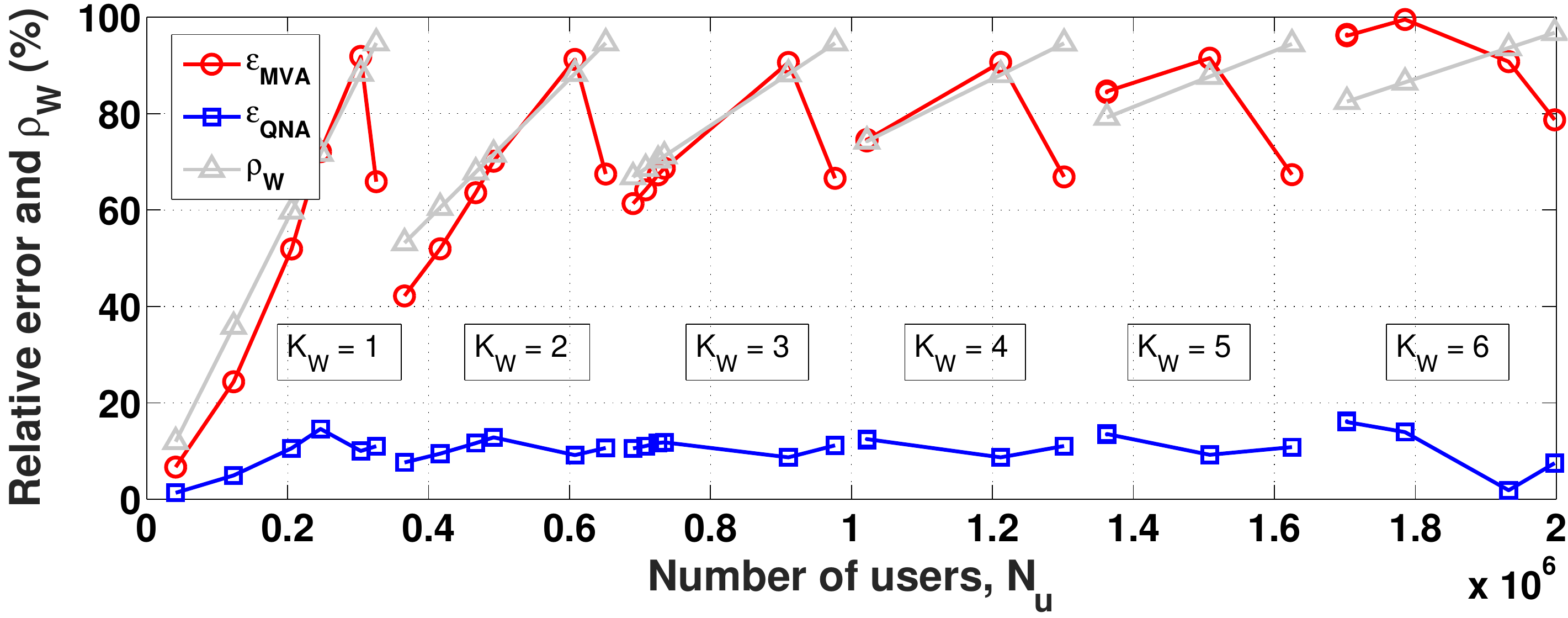}
\caption{Model validation.}
\label{subfig:relative-error}
\end{figure}

We computed the mean response time versus the number of users $N_{U}$ (see Fig. \ref{subfig:mean-response-time}).

When $T$ equals the target mean response time, $T_{target} = 3 ms$, a new W instance with a single CPU core is added.

For $N_{U} = 2 \cdot 10^6$, the FE utilization $\rho_{FE} = 0.95$ and from this point, the FE should be scaled to verify that $T \leq T_{target}$. Noteworthy, Jackson and MVA methods give similar results for the case studied.

From the data showed in Fig. \ref{subfig:mean-response-time}, we computed the relative error for the different analytical models considered as $\epsilon = \frac{|T_{sim} - T_{theo}|}{T_{sim}}$. In this case, $T_{sim}$ and $T_{theo}$ are respectively the mean response time obtained by simulation and computed by using the corresponding model.

We observed that our model is more accurate than those considered as baseline approaches (see Fig. \ref{subfig:relative-error}).

\section{Conclusions and Future Work}
\label{sec:conclusions}
In this paper, we propose an analytical model based on an open queuing network to estimate the mean response time of a VNF with several VNFCs. This model can be easily extended to be applied to chains of VNFs. The proposed model is sufficiently general to capture the complex behavior of such systems. 
To analyze the queuing network, we adapt the methodology proposed in \cite{whitt83} which is an approximate method to derive the performance metrics of a network of G/G/m queues. 

We validate our model by simulation. Specifically,
we validate it for an LTE virtualized Mobility Management Entity with a three-tiered architecture use case.
We also compare our model with 
the estate of the art, in terms of computational complexity and estimation error.
The results show that our model has a computational complexity 
similar to the method for analyzing Jackson's networks. Additionally, our model exhibits an estimation error, measured as the relative error for the estimation of the mean response time, approximately equal to 10\%, whereas for the considered baseline systems it ranges roughly from 60\% to 90\%.

Regarding the future work, several challenges lie ahead. One of the main challenges is to conduct a field evaluation of the proposed model to carry out a stronger validation of it.
Another challenge is to extend the model to make it more generic. On the one hand, the model should be able to capture the performance degradation due to the sharing of physical machine resources by the different VMs running on it. On the other hand, other virtualization scenarios should be considered, like VNFs implemented using microservices architectures.




\section*{Acknowledgement}

This work is partially supported by the Spanish Ministry of Economy and Competitiveness and the European Regional Development Fund (Projects TIN2013-46223-P, and TEC2016-76795-C6-4-R), and the Spanish Ministry of Education, Culture and Sport (FPU Grant 13/04833).

%
\bibliographystyle{IEEEtran}
\bibliography{IEEEabrv,references}

\end{document}